\documentclass[superscriptaddress, twocolumn,amslatex]{revtex4-1}

\usepackage{natbib}
\usepackage{amsmath}
\usepackage{tikz}
\usetikzlibrary{arrows,positioning}
\usepackage{mathtools}

\usepackage{verbatim}
\usepackage{color}
\usepackage{amsfonts}
\usepackage{graphicx}
\usepackage{dcolumn}
\usepackage{bm}

\include{partition}

\newcommand{\ket}[1]{\vert#1\rangle}

\begin{document}

\title{Error correcting power of small topological codes}

\author{Naomi H. Nickerson}
\affiliation{Department of Physics, Imperial College London, Prince Consort Road, London SW7 2AZ, UK } 
\email{n.nickerson11@imperial.ac.uk}

\begin{abstract}

Many quantum technologies are now reaching a high level of maturity and control, and it is likely that the first demonstrations of suppression of naturally occurring quantum noise using small topological error correcting codes will soon be possible. In doing this it will be necessary to define clear achievable metrics to measure and compare the performance of real codes. Here we consider the smallest examples of several families of topological codes: surface codes, color codes, and the gauge color code, and determine the minimum requirements to demonstrate error suppression. We use an exact decoder to compare the performance of the codes under simple models of both physical error and measurement error, and determine regions of correctability in the physical parameters for each code.

\end{abstract}

\maketitle

\section{Introduction}

There has recently been great progress in the level of quantum control in a number of physical qubit implementations. Trapped ions can be manipulated with error rates below $0.1\%$~\cite{Harty2014,Ballance2016} for all operations, and a fully programmable five qubit device has been developed~\cite{Debnath2016}. Superconducting qubits have demonstrated operations below $0.7\%$~\cite{Barends2014,Jeffrey2014}, and NV centre qubits can be controlled with errors around the $1\%$ level~\cite{Dolde2014}. These error rates place these systems either at, or in some cases well below the thresholds needed for fault-tolerant computation using the surface code~\cite{Wang2011}. Many experimental systems are thus poised to realise the first demonstrations of quantum error correction.

Furthermore, several recent experiments have demonstrated proof-of-principle error correction results in trapped ions~\cite{Nigg2014a}, superconducting qubits~\cite{Corcoles2015,Gladchenko2008,Kelly2015,Reed2012a}, NV centres~\cite{Cramer2015}, and solid-state spin systems~\cite{Waldherr2014a}. However, so far no physical realisation has, to our knowledge, demonstrated full error suppression, of both bit-flip and phase errors, where the rate of error on the encoded logical qubit is smaller than the error on a single, unprotected qubit under naturally occurring environmental noise. As technologies rapidly progress, this initial goal will likely soon become a possibility. Here we aim to identify and compare the minimum resource and error requirements to demonstrate error suppression for several families of topological codes: surface codes, color codes, and gauge color codes.

\begin{figure}[t]
\includegraphics[width=\columnwidth]{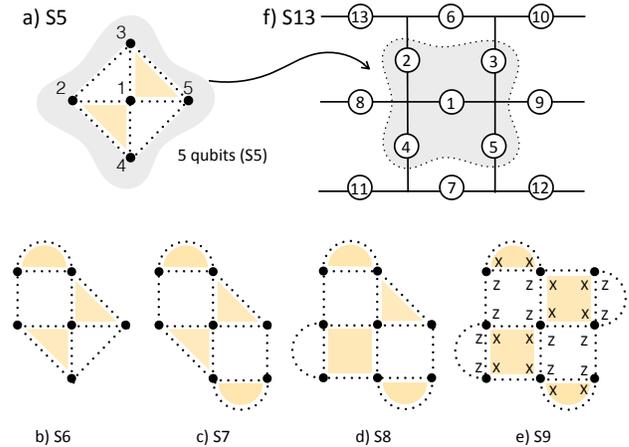}
  \caption[Small surface codes of between 5 and 13 qubits.]{Small surface codes of between 5 and 13 qubits, we label the surface code with $n$ qubits as S$n$. We show the codes in two representations (b) The most common lattice representation of the 13 qubit planar code, $X$ (star) stabilizers are defined at each vertex in the lattice and $Z$ (plaquette) operators on each face. The qubits are numbered to show in which order they are added as we increase the size of the code. In (a) and (c)-(f) codes of between 5 and 9 qubits are shown on a rotated lattice, $X$ stabilizers are defined on every colored face, and $Z$ stabilizers on every white face. The equivalence of the two representations is shown for the 5-qubit case (top). }
\label{fig:SC_surface_codes}
\end{figure}

Topological codes are the leading candidates for error correction on a large scale, they offer high thresholds, and require only low-weight, local operations. Of the topological codes, the surface codes~\cite{Kitaev2003} have come to dominate proposals for implementations of scalable quantum computing as they offer by far the highest thresholds under realistic noise models~\cite{Wang2011,Fowler2009}. Moreover, they require the smallest weight (4-body) measurements, which will be experimentally easier to perform, and less likely to induce errors than higher weight operators. However, when {\em computation}, rather than only quantum {\em memory} is considered, the color codes and gauge color codes offer simpler implementations of logical gates, reducing the qubit overhead to perform a desired algorithm. Thus there is a tradeoff between these properties - ease of implementation, error tolerance, and the number of qubits required to achieve a given level of error suppression. 

Here we consider the smallest possible demonstrations of error correction for each of these codes. Importantly, since topological codes are based on local, low-weight operations, these small scale demonstrations include all the operations needed to build a large scale code. The performance of small surface codes has also been discussed in ~\cite{Tomita2014} and~\cite{Wootton2016}. Here we also consider the performance of small color codes, and a small gauge color code. 

We remark that there are a number of other codes which may also have the potential to demonstrate error suppression with small numbers of qubits, but that do not have the same potential for scalability as the topological codes that we study here. The smallest code capable of correcting all single qubit errors is the 5-qubit code~\cite{Laflamme1996}. There have been recent proposals for the implementation of this code in trapped ions~\cite{Goodwin2014b}. The Bacon-Shor code~\cite{Bacon2006,Shor1995}, as another example, does not have a threshold, but may well offer good error suppression especially under asymmetric noise. 

\bigskip

The performance of error correcting codes is usually understood in terms of the {\em threshold} of the code, the physical error rate below which the logical failure rate of the code can be arbitrarily suppressed by increasing the code size. These thresholds are relevant quantities when one considers large code dimensions, but here we consider codes of a fixed size. In order to assess the performance of small codes we require a different metric 

 To quantify the performance of the codes we will instead determine their {\em correcting power}, which we define to be the ratio of the logical failure rate, $p_L$ to the physical failure rate, $p$, $\mathcal{C} = p/p_L $. We say a code has {\em error correcting power} if the logical error rate is smaller than the physical error rate, $\mathcal{C}>1$.

\bigskip

The remainder of this manuscript is structured as follows: in Section~\ref{sec:SC_codes} we introduce and define the codes we will consider and discuss their relative merits in terms of fault tolerant operations and scalability. In Section~\ref{sec:SC_error_model} we introduce several simple error models, and our approach to decoding. In Sections~\ref{sec:SC_perfect_measurement} and~\ref{sec:SC_noisy_measurement} we present the results of simulations determining the correcting power of each code under various error scenarios and determine the conditions under which error suppression can be demonstrated.

\section{Error correcting codes}

\label{sec:SC_codes}

The practical considerations of a high error threshold, and simplicity of implementation favour the surface code for experimental fault-tolerance, and these are indeed important characteristics of any code. But with a view to going beyond a fault tolerant quantum {\em memory}, towards implementing fault tolerant quantum {\em computation}, there are other properties of the code that also become important. Namely: how easy will it be to perform protected logical operations? For ease of implementation {\em transversal gates} are highly desirable, as they do not require any significant additional overhead, and do not propagate errors. A gate is transversal if the logical gate can be performed by applying only single qubit rotations to each physical qubit in the code, or for a logical two-qubit gate between two encoded states, applying two-qubit gates to pairs of physical qubits between the two codes~\cite{NielsenChuang}. Logical operations that cannot be performed transversally require the introduction of additional resources, such as {\em magic states}~\cite{Bravyi2005}, which incur a significant additional cost in both the number of physical qubits, and the time taken for computation. Different codes support different sets of transversal operations which will dramatically change the overheads required to implement logical gates.  

We consider several classes of codes, which can be understood via their stabilizer generators, $S_i$, which we will refer to as the {\em stabilizers} of the code. The codespace is the simultaneous eigenspace of the stabilizers such that $S_{i} \ket{\psi_c} = \ket{\psi_c}$, where $\ket{\psi_c}$ is a state in the codespace. 

Let us briefly review the procedure for error correction in a stabilizer code. An initially perfect code first acquires some Pauli error. It is sufficient to consider only Pauli-$X$ errors, which we refer to as {\em bit-flip errors} and Pauli-$Z$ errors, which we refer to as {\em phase errors}, since other forms of noise can be decomposed into errors in these two channels. The stabilizers are then measured in order to gather information on the location of errors. Pauli errors occurring on the qubits will anticommute with some of the stabilizers, causing their measurement outcomes to be flipped to their `-1' eigenvalue. Measuring all the stabilizers returns a {\em syndrome} of classical information. The locations of `-1' outcomes in the syndrome can be used to identify and correct for the presence of errors. 

In all the codes we consider the stabilizers are essentially parity checks on groups of qubits, in either the $X$ or $Z$ basis. In practice the most straightforward approach to stabilizer measurement requires an ancilla, or {\em measurement qubit} which will interact directly with each qubit in the stabilizer before being measured out to reveal the result of the parity check. We do not include these measurement qubits in our description of the code. In reality at least one additional measurement qubit would be required in addition to the numbers we state later to implement error correction, which could be used to measure each of the stabilizers sequentially. Alternatively many measurement qubits could be dispersed throughout the lattice, one for each stabilizer.

\begin{figure}
\includegraphics[width=\columnwidth]{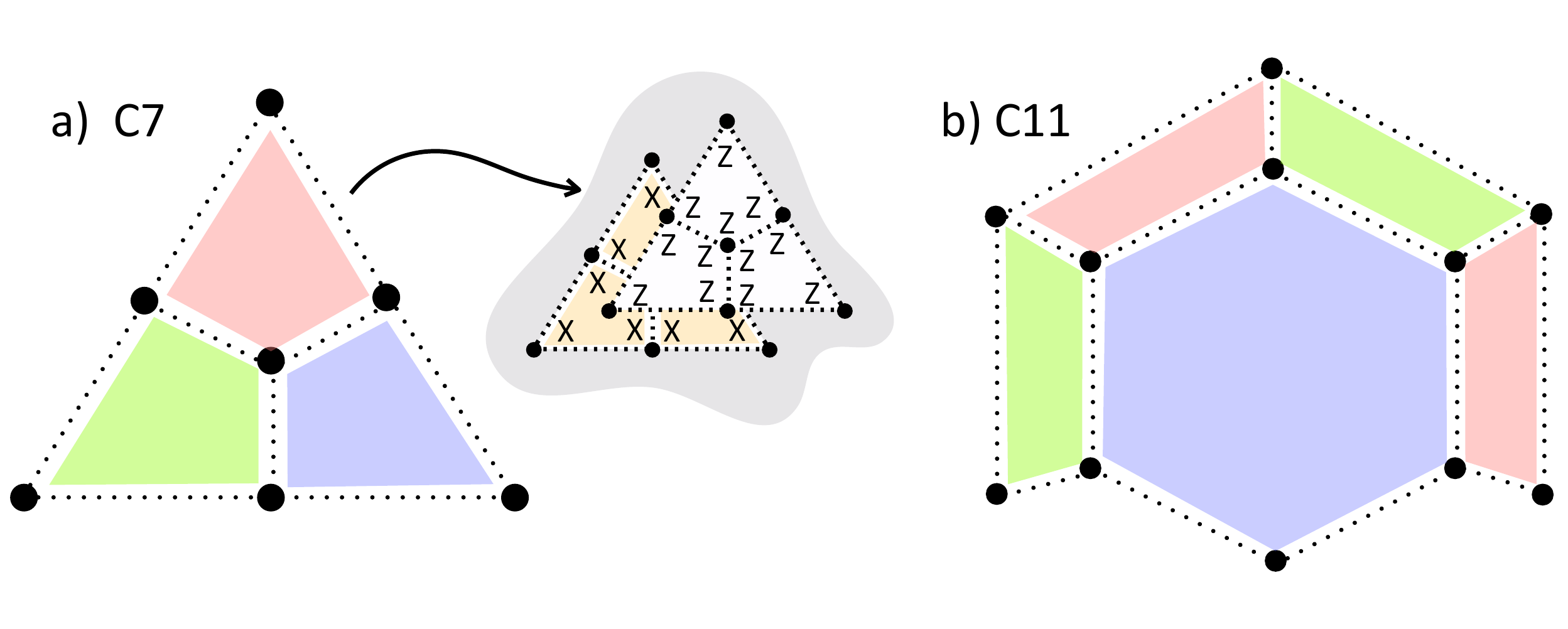}
\caption[Small color codes]{(a) 7 qubit color code (C7). The code places qubits at the centre, edges and vertices of a triangle. Stabilizers are defined on the faces of the lattice. For each face there is an $X$-type and a $Z$-type stabilizer. Unlike in the surface codes these stabilizers fully overlap, and representing the code with the coloring scheme of Figure~\ref{fig:SC_surface_codes} we can view this as two overlaid triangles (inset). (b) An 11 qubit color code (C11), which comprises two 6-body stabilizers and eight 4-body stabilizers. }
\label{color_codes}
\end{figure}

\subsection{The surface codes}
\label{sec:SC_surface_codes}

We consider the planar variant of Kitaev's surface code~\cite{Kitaev2003}, this is a 2D stabilizer code defined on a square lattice. The surface code has the highest known error threshold of any topological code, with a threshold error rate of $\sim$10.3\% under random noise (in each error channel) ~\cite{Wang2003}, which falls to around 1\% when considering a noisy circuit-level implementation~\cite{Wang2011}. These thresholds are summarised in Table~\ref{tab:code_comparison}. Surface codes are also appealing because of the low-weight measurements required, with only 3 or 4 qubits in each stabilizer, this contributes to the high thresholds and also makes physical implementation easier. 

In the surface code, the logical Pauli operations and the logical {\small CNOT} gate can be performed transversally, while a Hadamard can be performed transversally only up to a physical rotation of the code. In order to implement a full universal gate set the surface code must be supplemented with magic states~\cite{Bravyi2005}, which, in combination with the code's transversal gates admits a logical $\frac{\pi}{8}$ gate. 

In the most common representation the qubits are placed on the edges of a square lattice as shown in Figure~\ref{fig:SC_surface_codes}(b). Here for clarity we consider a rotated representation of the code where qubits reside on the vertices of a square lattice, and stabilizers are defined on each face. In this representation the $Z$-type operators (plaquettes) and $X$-type operators reside on alternating faces as shown in Figure~\ref{fig:SC_surface_codes}. We consider surface codes of 5-9 and 13 qubits, which we will refer to as as S5-S13; these are shown in Figure~\ref{fig:SC_surface_codes}(a)-(f). 

In the usual square lattice configuration S5 and S13 are the two smallest variants of the planar code. S5 can correct only $Y$ errors, while S13 is the smallest capable of identifying all single qubit errors. One can, however, also consider a rotated variant of the code ~\cite{Bombin2007a,Horsman2012,Tomita2014}. Removing the 4 corner qubits from S13 results in a 9 qubit code, S9, (see Figure~\ref{fig:SC_surface_codes}(f)) which is also capable of correcting any single qubit error. It is worth noting that this 9 qubit variant cannot uniquely {\em identify} all single qubit errors, as there are pairs of errors at the boundary of the code which produce identical syndromes. However, these pairs form the weight-2 stabilizers at code edges, meaning an incorrect `guess' at the error configuration will simply result in a stabilizer acting on the code, leaving the code space undisturbed. 

\label{sec:S8_ec}
We remark that although S9 is the smallest code that can correct all single qubit errors, the eight qubit code (S8) is the smallest capable of demonstrating {\em error suppression}. This code can correct all single qubit Pauli-$Z$ errors, and all but one Pauli-$X$ error. Thus in the limit $p \rightarrow 0$ we find a logical error rate of $p_L = p_x$, while the single qubit error rate is $p \sim p_x + p_z$. Furthermore, other variants of the code are able to suppress errors under certain error models. The seven qubit surface code (S7) shows some limited error suppression under depolarising noise, and under the very restrictive case of only Pauli-$Y$ errors, even the five qubit surface code can also show corrective power.

Finally, we note that S6-8 are not symmetric in their $X$ and $Z$ stabilizers, and consequently their correcting power in these two channels will differ. In many physical systems it is indeed the case that the rate of $X$-type (bit-flip) errors is different to the rate of $Z$-type (phase) errors. This asymmetry of the codes can then act to our advantage. With a large enough discrepancy we will find that S7 and S8 can demonstrate high levels of error suppression. More generally, the surface codes have no requirement on symmetry between the plaquette and star operators, leaving us the ability to shape a code to give greater protection against one channel should it be required.


\begin{table*}[t]

\centering
\small
 
\begin{tabular}{ |p{2cm}||p{3.6cm}|p{4.3cm}|p{3.7cm}|  }

 \hline
   & \multicolumn{3}{|c|}{Thresholds}   \\
 \hline
    Code  &  Physical errors (i.i.d) & Measurement and physical \newline errors  & Circuit level noise \\
 \hline
 Surface Code \newline  & 10.31\% ~\cite{Wang2003}  &  2.9\% ~\cite{Wang2003} &   1.1\%~\cite{Wang2011}  \\
 \hline
 Color Code \newline &   10.6\%~\cite{Landahl2011}   & 3.1\%~\cite{Landahl2011}  & 0.082\%~\cite{Landahl2011} \\
 \hline
 Gauge Color \newline Code & 0.46\%~\cite{Brown2015} \newline & 0.31\%~\cite{Brown2015} &  - \\

 \hline
\end{tabular}

\bigskip

 \begin{tabular}{ |p{2cm}||p{0.9cm}|p{2cm}|p{3.5cm}|p{1.4cm}|p{3.5cm}|  }

 \hline
   & \multicolumn{5}{|c|}{Code Properties}   \\
 \hline
    Code  & Dim. &Min. qubits \newline for error \newline correction & Transversal gates & Requires magic states? & Measurement \newline error correction   \\
 \hline
 \hline
 Surface Code & 2D & 9 $^a$ &  {\small X, Z, CNOT} \newline (+ {\small H} $^b$ )  & Yes & $\sim{L}$ rounds of stabilizer \newline measurement $^c$    \\
 \hline
 Color Code & 2D & 7 & {\small X, Z, CNOT, H, S} \newline (full Clifford group) & Yes & $\sim{L}$ rounds of stabilizer\newline measurement  \\
 \hline
 Gauge Color \newline Code & 3D & 15 & {\small X, Z, CNOT, H, S, $\pi/8$}\newline (Universal operations) & No & Single-shot \newline error correction\\

 \hline
\end{tabular}

  \caption{
  A comparison of the three families of codes we study in this manuscript. The upper table shows the thresholds of each code under three noise models. Physical errors, in which random i.i.d noise is the only noise suffered by the qubits. Measurement and physical errors, also known as {\em phenomenological noise}, where qubits suffer physical i.i.d noise with some probability, $p$, and additionally measurements are faulty with the same probability. The final column indicates the threshold under a full circuit noise model, where all basic operations are considered to be noisy. The lower table summarises the main characteristics of each code type.  \newline 
  \footnotesize{
  a. 9 qubits is the smallest number needed to allow correction of all single qubit errors, however, a smaller number of qubits may be sufficient to demonstrate error suppression under certain noise models, see Section~\ref{sec:S8_ec} \newline
	b. The Hadamard gate is transversal in the surface code up to a physical rotation of the code \newline 
      c. Where $L$ is the physical dimension of the code}
  }
 
  \label{tab:code_comparison}

\end{table*}


\subsection{Color codes}
	
The second family of topological codes that we consider are the color codes~\cite{Bombin2006}. Like the surface codes, these are 2D stabilizer codes. From a computational perspective the color codes offer an advantage over the surface codes in that the logical Hadamard gate, $H$, and the logical phase gate, $S$, can be performed transversally~\cite{Bombin2006}, thus lowering the overheads for computation. 

The main disadvantage of the color codes, in comparison to the surface codes, is their lower error thresholds. While their performance under random noise is high (see Table~\ref{tab:code_comparison}), currently the best threshold achieved under full circuit noise is 0.143\%~\cite{Stephens2014}, 

an order of magnitude lower than that of the surface code. Despite this, there is some evidence that in the far-below-threshold limit the color codes may still be more efficient than surface codes in the number of qubits required to achieve a given logical error rate~\cite{Landahl2014}. Another possible drawback of the color codes is the higher-body measurements they require, with either 6- or 8-body stabilizers needed for larger versions of the codes. This feature may make an experimental implementation more challenging, and result in larger measurement errors. 

The color codes are defined on a trivalent lattice, where qubits lie on the vertices, as shown for two small examples in Figure~\ref{color_codes}. For each face of the lattice, $f$, there are two associated stabilizer generators: $X_f$ which is the product of Pauli-$X$ operators acting of every qubit on the boundary of the face, and $Z_f$, which is the product of Pauli-$Z$ operators on the same set of qubits. 
It is worth noting that this construction means that the color codes are inherently symmetric between the X and Z bases, and so the color codes will always be at their most efficient when $p_x = p_z$. Unlike the surface codes, there is no mechanism by which the code can be adapted to imbalances in these error rates.

We consider two small color codes, which are shown in Figure ~\ref{color_codes}. The 7 qubit color code (C7) is the smallest possible color code, and is capable of correcting all single qubit errors.

\subsection{Gauge color codes}

\begin{figure}
\centering
\includegraphics[width=\columnwidth]{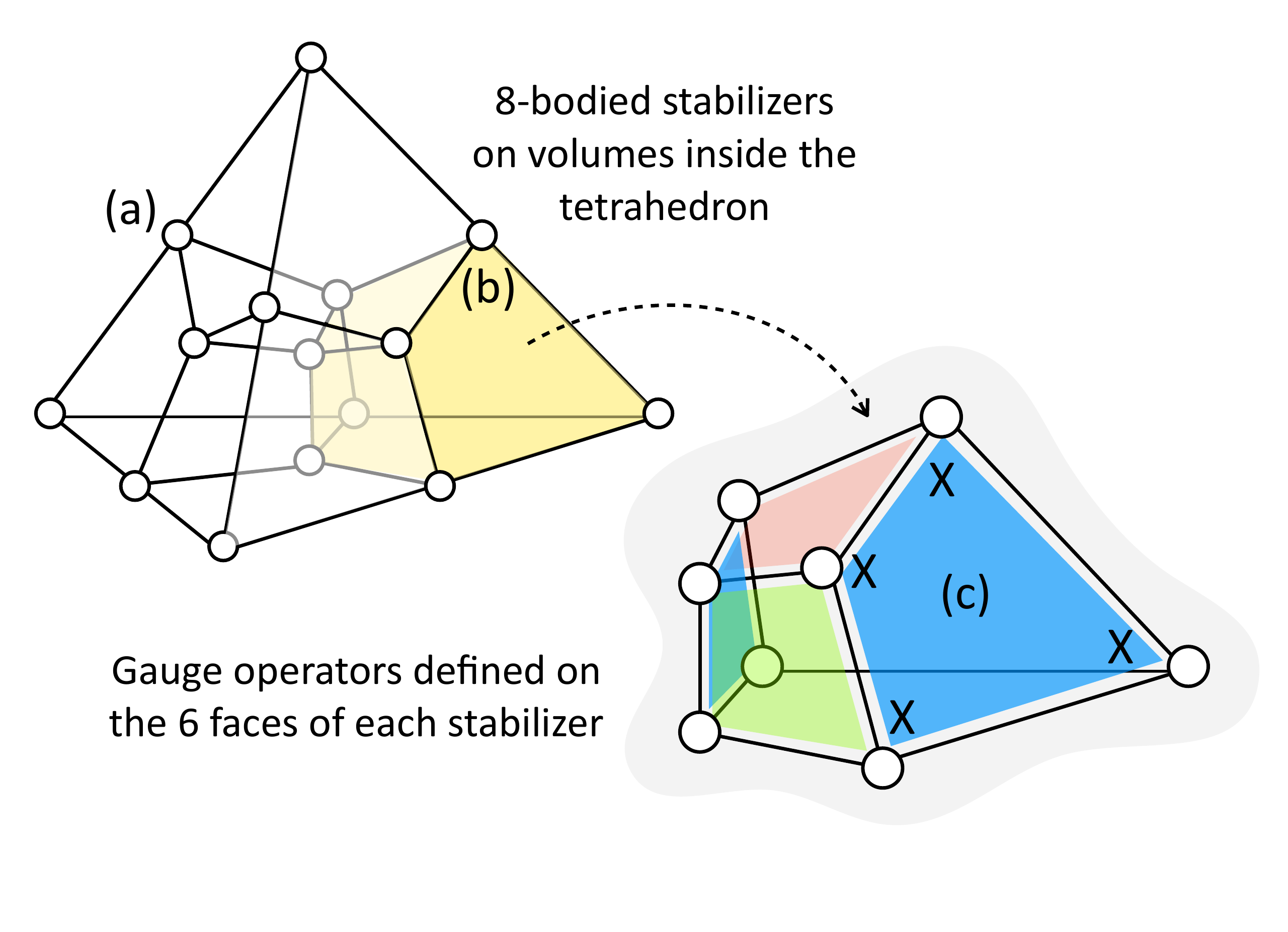}
\centering
\caption[The 15 qubit gauge color code.]{\textbf{The 15 qubit gauge color code.} (a) Qubits are represented by white circles that reside in the body, and on the faces, edges, and vertices of a tetrahedron (b) One stabilizer volume is highlighted in yellow. The 8 stabilizers are defined as $X^{\otimes 8}$ and $Z^{\otimes 8}$ operators on each volume inside the tetrahedron (c) One $X^{\otimes 4}$ gauge operator is represented on a blue face of a stabilizer cell. The 36 gauge operators are defined as $X^{\otimes 4}$ and $Z^{\otimes 4}$ for each face in the lattice. } 
\label{15qubit_gcc}
\end{figure}

Finally we consider the 3D gauge color codes, recently proposed by Bombin~\cite{Bombin2013}, which are topological subsystem codes. This type of code is of particular interest as it offers both error correcting and computational properties beyond that of the surface codes. Firstly, it has been shown~\cite{BombinSingleShot,Brown2014} that {\em single-shot fault tolerance} is possible with the gauge color code. That is, some level of measurement error can be tolerated with only a {\em single round} of stabilizer measurement. For the surface codes, and color codes, tolerance to measurement error can only be achieved by making many rounds of stabilizer measurement, requiring a far greater time and level of operational complexity. 
The second advantage of the gauge color codes is that they admit a universal transversal gate set without needing to resort to magic state distillation~\cite{Kubica2014}. 

We consider the smallest gauge color code, which is shown in Figure~\ref{15qubit_gcc}. This code comprises 15 qubits, which are positioned on a tetrahedron, one at the centre of the volume, and one on each face, edge and vertex. This encodes a single logical qubit. For each {\em internal volume} of the tetrahedra, $v$, (as shown in Figure~\ref{15qubit_gcc}), there are two associated stabilizer generators: $X_v$ which is the product of Pauli-$X$ operators acting on every qubit on the boundary of the volume, and $Z_v$, which is the product of Pauli-$Z$ operators on the same set of qubits. 

The reader will note that there 8 stabilizer generators, for a code made up of 15 qubits and encoding a single logical qubit. The remaining 6 degrees of freedom make up the {\em gauge}. For each face, $f$, in the lattice there is a gauge operator, $G^{X}_{f}$, that is the tensor product of Pauli-$X$ operators acting on each qubit	 on the vertices of the face, and another gauge operator, $G^{Z}_{f}$, that is the tensor product of Pauli-$Z$ operator acting on the same set of qubits. In total there are 36 gauge operators. 

An important feature, from a practical perspective, is that physically we are not required to measure the stabilizers themselves, but instead need only measure the smaller-bodied gauge operators, from which the stabilizer information can be reconstructed. Each stabilizer can be decomposed into three distinct pairs of gauge operators. This can be seen by observing that the surface of each stabilizer is three-colorable, as shown in Figure~\ref{15qubit_gcc}. The two blue faces, for example, cover all the qubits in the volume, and so the product of the two corresponding gauge operators is equal to the stabilizer of that volume. The same is equivalently true for the red and green pairs of faces. When the gauge operators are measured without error the values of the three pairs should agree. If they do not, we can use this redundancy to identify measurement errors.

Finally, we note that this code is also the smallest possible construction of the recently proposed {\em doubled color code}~\cite{Bravyi2015}, or {\em stacked color code}~\cite{Connor2015}. These codes remarkably can achieve a full transversal gate set whilst maintaining a 2-dimensional code structure. However their fault tolerance properties in larger lattice sizes have not yet been fully understood.

\section{Error Model}
\label{sec:SC_error_model}

We distinguish between two types of error that can occur during an operation: {\em physical errors} and {\em measurement errors}. We consider two models for physical errors: {\em depolarising noise}, and {\em independent noise}, in which $X$ and $Z$ type errors are uncorrelated. In both error models we normalise the parameters to be described by $p$, the probability with which a single qubit acquires any error. In both cases we treat each qubit as independent. Both of these physical error models will be considered under perfect measurement, and the more realistic case of noisy measurements. 

\subsection{Physical errors}

\paragraph{Depolarising noise}
Each qubit has a probability $p$ of acquiring some error which is equally likely to be one of the three Pauli errors, such that 
\[ p_x = p_y = p_z = p/3. \]

\paragraph{Independent $X$ and $Z$ errors}
Each qubit has a probability $p'_x$ of acquiring a Pauli-$X$ error and $p'_z$ of a Pauli-$Z$ error, where we introduce the prime notation in this case to indicate that these are different physical quantities to the error rates under depolarising noise. The overall probability that a qubit acquires an error is
\[ p = 1 - (1-p'_x)(1-p'_z). \]
To consider the performance of the various codes under different ratios of errors between the two channels we describe the relationship between the two as, 
\[ p'_z = \alpha p'_x .\]
We will study first the special case of $\alpha = 1$, where there is no imbalance between the channels, and then consider the more general case of unequally distributed $X$ and $Z$ errors. We will consider explicitly here only cases of $\alpha>1$, that is, where phase errors are more likely to occur than bit-flip errors, since this is the more common scenario in physical systems. 

\subsection{Measurement errors}
We define the measurement error rate, $q$, to be the probability that a single stabilizer measurement misreports, and returns the opposite result when it is evaluated.  

\subsection{Error correcting power} 

We will characterise the codes by calculating their {\em error correcting power}, $\mathcal{C}$, which we define as the ratio of the error rate on a single qubit to the logical error rate after correction. Here we will only consider a single round of stabilizer measurement and correction, and assume that the initial encoded state can be perfectly prepared. In the case of perfect measurement, 
\[ \mathcal{C} = p/p_L. \]
A value of $\mathcal{C}>1$ indicates a suppression of the error under encoding.

In the case of measurement error then we must consider the error rate of a single unencoded qubit to be the probability that neither a physical error, nor a measurement error occurs during storage and readout.  The logical error rate of the encoded qubit accounts for measurement error during the decoding process. In this case the correcting power is given by,
\[ \mathcal{C} = (1- (1-p)(1-q))/p_L. \]
%


\section{Decoding}
\label{sec:SC_decoding}

The procedure for error correction involves measurement of the stabilizers of the code. These parity checks should all return a `+1' outcome if the qubits are error free. If an error has occurred, however, this anticommutes with some of the stabilizers and flips their measurement outcome to `-1'. The results of all the stabilizer measurements is named the {\em syndrome}. The task of deducing, from the syndrome, the best possible way to try and fix the code is the task of the {\em decoder}.  

For high-distance codes approximate methods are used to reduce the computational requirements of decoding~\cite{Fowler2013c,WoottonLoss,Anwar2014,Hutter2014d,Stephens2014a,Brown2015}. For surface codes in particular these have been heavily optimised and can achieve close to optimal thresholds~\cite{Fowler2012g,Hutter2014}. Here, however, since we consider only small system sizes, it is possible to use an exact decoder to identify an optimal decoding under both physical errors, and measurement errors. In the following we use a precomputed decoder~\cite{TomThesis} to compile a lookup-table of syndromes and their corresponding error configurations, which can then be used to deduce the optimal performance of the codes. We describe the decoder in detail in Appendix~\ref{appendix:decoder}.

\section{Perfect measurement}
\label{sec:SC_perfect_measurement}

\begin{figure}
\begin{center}
\includegraphics[width=\columnwidth]{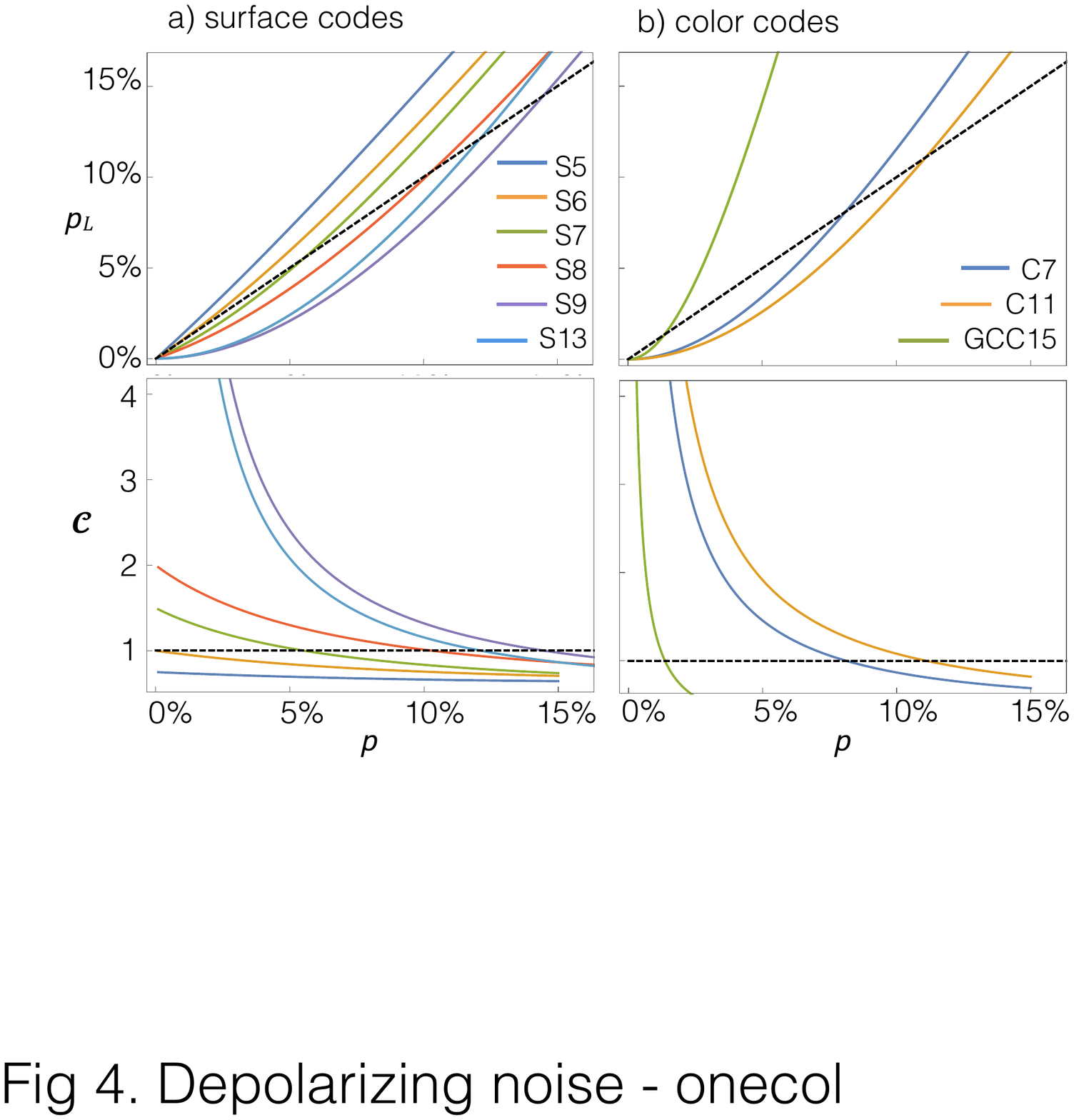}
\end{center}
\caption[Small codes under depolarising noise and perfect measurement.]{\textbf{Depolarising noise and perfect measurement.} Logical error rates (upper plots) and error correcting power (lower plots) under depolarising noise and perfect measurement for (a) surface codes and (b) color codes. The single qubit error rate is shown for comparison (black dashed line). In the lower plots, the region of the curves lying above the dashed line indicate the code has error correcting power, as the logical error rate is suppressed relative to the physical error rate. Amongst the surface codes S7 and S8 are the smallest to offer error correcting power $>1$. These two codes reach a finite maximum $\mathcal{C}$ as $p\rightarrow0$ as they are not capable of correcting all possible single qubit errors. S9 and S13 show divergent error correcting power as $p\rightarrow0$.}
\label{depol_plots}
\end{figure}

We first study the codes under perfect measurement, using the exact decoder to calculate their correcting power, $\mathcal{C}$, as a function of the physical error rate, $p$. 

{\bf Depolarising noise}. The performance of the codes under depolarising noise is shown in Figure~\ref{depol_plots}. Amongst the surface codes, S7 is the smallest to demonstrate error correction, with $\mathcal{C}>1$ for a physical error rate below $\sim 5 \%$. S8 shows error correction for $p < 10\%$. S7 and S8 cannot correct for all single qubit errors, and consequently their error correcting power reaches a finite maximum value as $p \rightarrow 0$. In S7, on 5 of the 7 qubits  all single qubit errors can be corrected, whereas of the remaining two, only $Z$ and $Y$ errors can be corrected. Thus in the limit that $p\rightarrow 0$, this results in an overall logical error rate of  $p_{L} = 2p_{x} = \frac{2}{3}p$, and so $\mathcal{C}_{\rm S7} \rightarrow 1.5$. In S8, any single qubit error can be detected and corrected on seven of the qubits   whilst one qubit can only correct $Z$ and $Y$ errors. Thus as $p\rightarrow 0$ we find $p_L \rightarrow p_x = \frac{p}{3}$. This is seen in Figure~\ref{depol_plots} where $\mathcal{C}_{\rm S8}\rightarrow 3$. S9 and S13, on the other hand, are able to correct all single qubit errors and so their correcting power diverges as $p\rightarrow 0$. S9 outperforms S13 as it offers the same code distance, but has fewer physical qubits.  Both color codes, and the gauge color code show a region of error suppression. C7 is correctable for $p<8\%$ and C11 for $p<11\%$. The gauge color code has a smaller region of correctability with $\mathcal{C}>1$ for $p<1.5\%$.

\begin{figure}
\begin{center}
\includegraphics[width=\columnwidth]{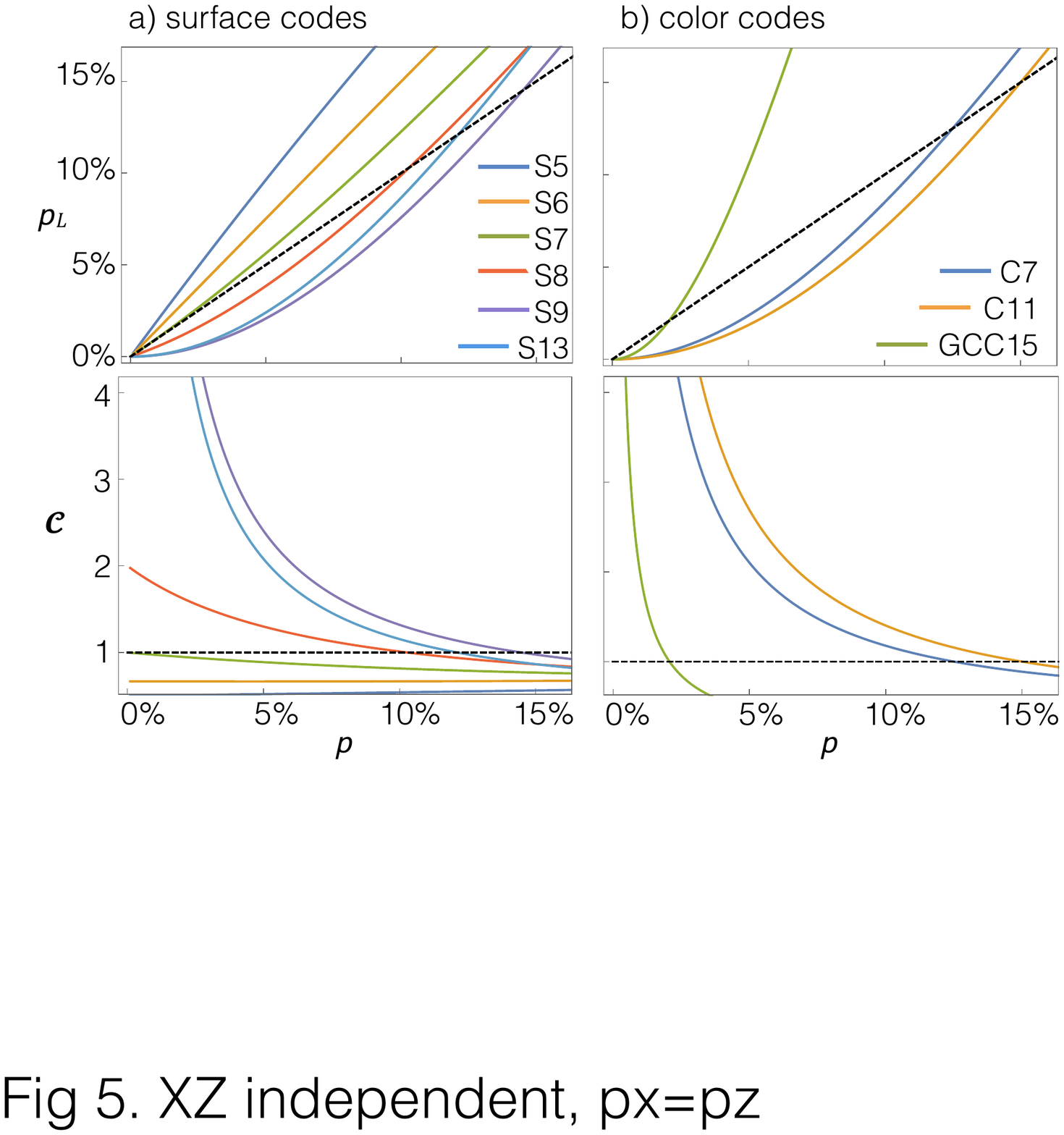}
\end{center}
\caption[Small codes under symmetric independent noise and perfect meausrement]{\textbf{Symmetric independent noise and perfect meausrement.} Logical error rates (upper plots) and error correcting power (lower plots) under symmetric independent noise ($p_x=p_z$) and perfect measurement for (a) surface codes and (b) color codes, as a function of the physical error rate, $p$. The single qubit success rate is shown for reference (black dashed line). For the surface codes the smallest code to demonstrate error suppression is the 8-qubit code. Both color codes and the gauge color code demonstrate error suppression.}
\label{xz_error_plot}
\end{figure}

{\bf Independent noise}. We first consider the symmetric case where $p'_x = p'_z$, the performance of the codes under this noise model are shown in Figure~\ref{xz_error_plot}. For the surface codes the performance is largely the same as under depolarising noise, with the exception of the 7-qubit code which no longer has any correcting power. The color codes, on the other hand, have greater correcting power than under depolarising noise. C7 and C9 are correctable for $p< 12.5\%$ and $p<15\%$ respectively, while the gauge color code is correctable below $p\sim 2\%$.

\begin{figure}
\begin{center}
\includegraphics[width=\columnwidth]{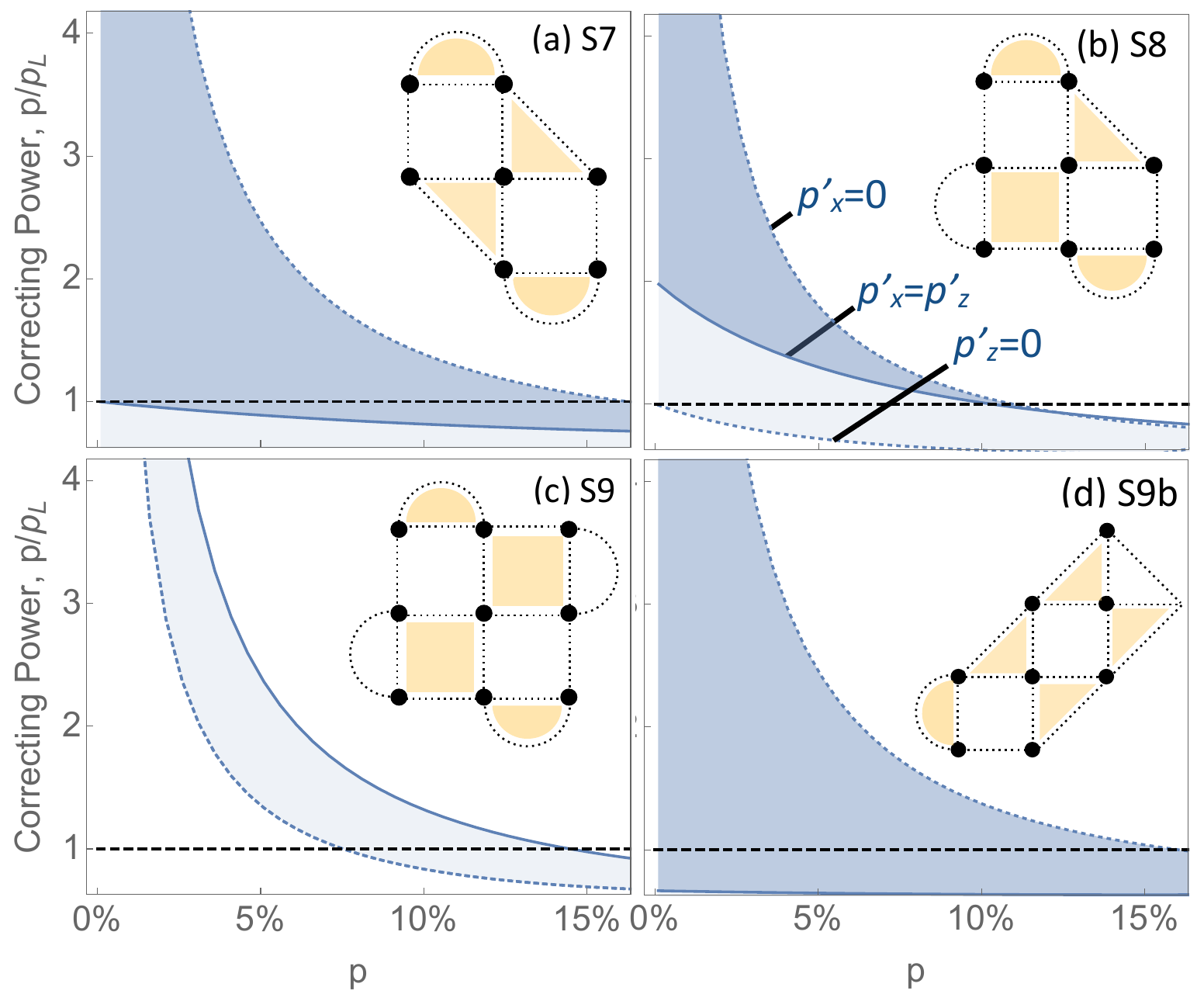}
\end{center}
\caption[Error correcting power of the surface codes under asymmetric noise]{{\bf Error correcting power of the surface codes under asymmetric noise.} Surface codes containing 7-9 qubits under independent noise, where in general $p'_x \neq p'_z$. The shaded region indicates the range of values the logical success rate can take for all possible $(p'_x,p'_z)$. $p$ is the single qubit error rate, $p = 1-(1-p'_x)(1-p'_z)$.  Curves are labelled in (b). The boundaries of the region, where $p'_x=0$ and $p'_z=0$ are shown as solid blue lines. The $p'_x = p'_z$ case is shown as a dotted blue line. The single qubit error rate, $p$, is shown for reference (dashed black line). The darker shaded region indicates a {\em reduced} logical error rate relative to the case of  $p'_x = p'_z$, whereas the lighter shaded region indicates an {\em increased} logical error rate. }
\label{xz_uneven_surface}
\end{figure}

We next consider the more general case where $p'_x\neq p'_z$. Codes that are symmetric between the $X$ and $Z$ bases will always perform worse in this scenario, as their rate of error suppression will be limited by the larger of the two error rates. Codes that have an asymmetry between the two bases, on the other hand, may show improved performance when a larger proportion of the errors are in the {\em preferred channel}. S8, for example, is capable of correcting all single qubit $Z$ errors, but only 7 out of 8 possible single qubit $X$ errors. Thus its performance improves when $p'_z > p'_x$.  The results for the logical error rates of several of the surface codes are summarised in Figure~\ref{xz_uneven_surface}. The case $p'_x = p'_z$ is indicated by the solid line, and the limiting cases of $p'_x = 0$ and $p'_z = 0$ are shown by the dotted lines. For the case of the symmetric code S9 (Figure~\ref{xz_uneven_surface}(c) we see that as the ratio of errors in the two channels changes the logical error rate increases.  The codes S7, S8 and S9b which have a structural asymmetry, however, show a reduced logical error rates when $p'_z > p'_x$. The darker shaded region indicates the area in which the logical error rate is reduced relative to the $p'_x = p'_z$ case. This effect is offset by a {\em reduced} performance when the imbalance is inverted, since the error correcting power has been concentrated in one channel. More detail on the performance under varying levels of asymmetry is given in Appendix~\ref{app:SC_asymmetry}.

\section{Noisy measurement}
\label{sec:SC_noisy_measurement}

\begin{figure}
\begin{center}
\includegraphics[width=\columnwidth]{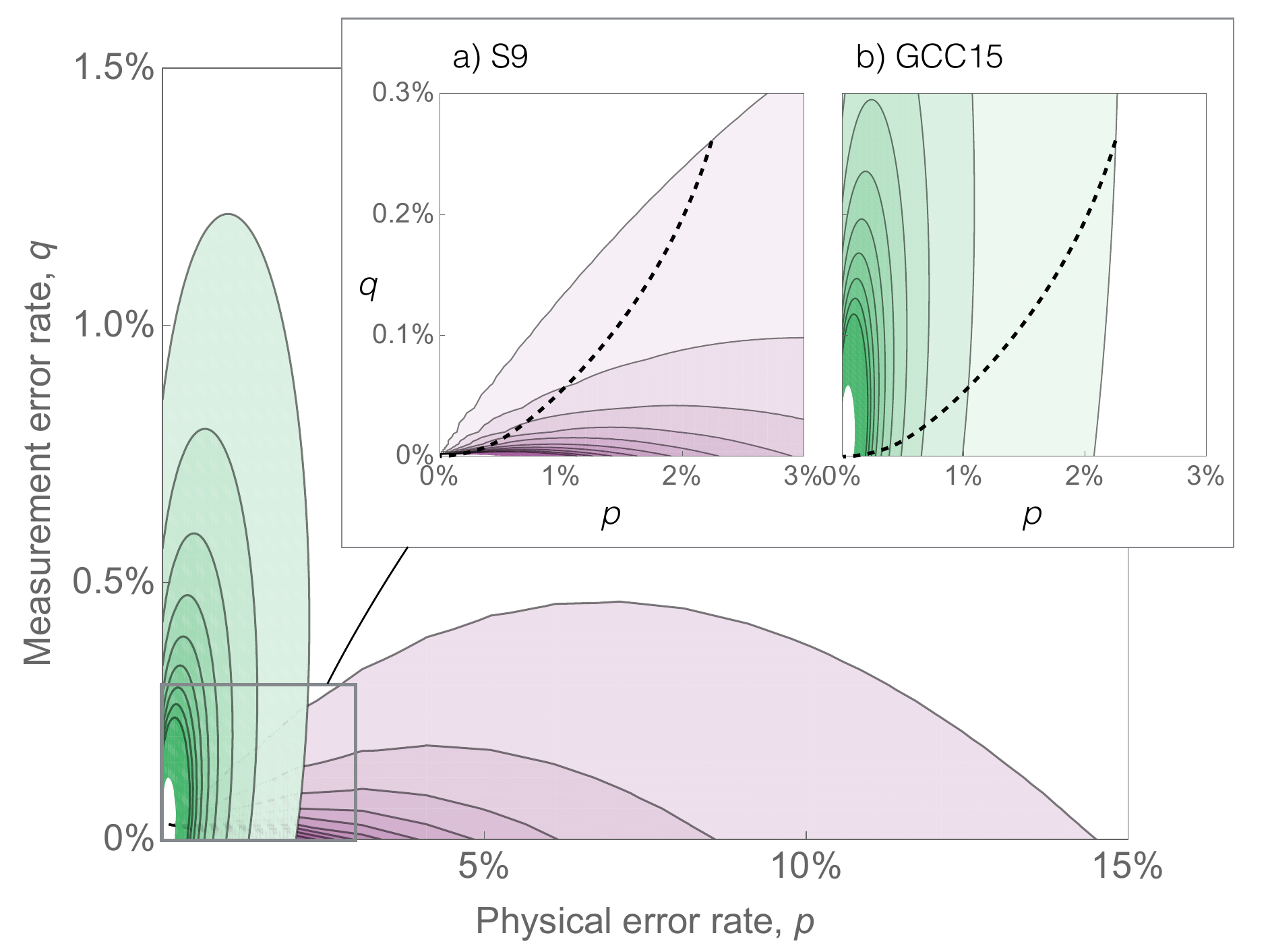}
\end{center}
\caption[Correcting power of small codes under noisy measurement]{\textbf{Noisy measurements and depolarising noise.} Correcting power of the S9 surface code (purple) and the 15 qubit gauge color code (green) under noisy measurement, and depolarising noise. The horizontal axis indicates the physical error rate, $p$, and the measurement error rate, $q$, is shown on the vertical axis. The shaded region in each plot indicates where $\mathcal{C}>1$, and thus the code is capable of demonstrating error suppression. The contours represent lines of constant error correcting power, $\mathcal{C}$. In the main figure contours indicate increments of 0.5, in the inset, increments are of 1. The inset shows a zoom on a) the 9 qubit surface code, and b) the 15 qubit gauge color code. At the black dashed line indicates the $(p,q)$ values where to two codes have equal error correcting power. }
\label{noisy_individual}
\end{figure}

The presence of measurement noise impacts heavily on the codes' performance. To achieve fault tolerance with a surface code or color code in this scenario it is necessary to perform multiple rounds of stabilizer measurement, but given low enough errors a single round of stabilizer evaluation can still suffice to demonstrate error suppression. Here we calculate the error correcting power of each code under a single round of stabilizer measurement to determine under what conditions this is true. Unlike the other codes, the gauge color code does allow the identification of some measurement errors. In the previous section we saw that under perfect measurement GCC15 simply performed more poorly than the other codes. However, we will now see that under certain combinations of physical error and measurement error it will significantly outperform the surface and color codes.

\begin{figure}[t]
\begin{center}
\includegraphics[width=\columnwidth]{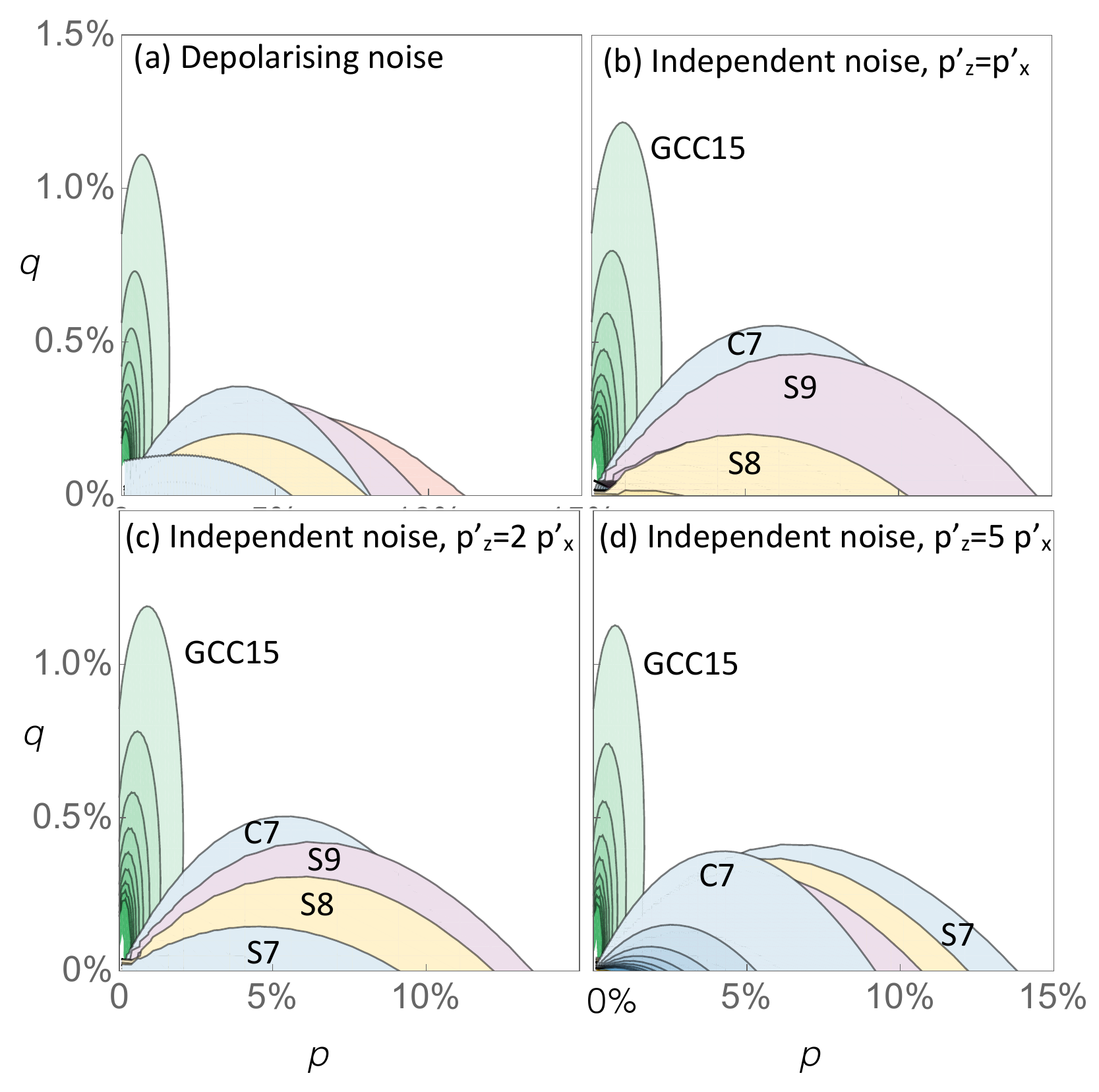}
\end{center}
\caption[Regions of correctability under noisy measurement]{\textbf{Regions of correctability under noisy measurement.} Regions of correctability in the space of measurement error, $q$, and physical error, $p$ are shown for surface codes of 7-13 qubits (solid lines), color codes of 7 and 9 qubits (dashed lines) and the 15 qubit gauge color code (dotted line). (a) Depolarising noise (b) symmetric independent noise, $p'_z=p'_x$, (c)-(d) asymmetric independent noise with $p'_z > p'_x$. }
\label{noisy_measurements}
\end{figure}

We identify a {\em region of correctability} in the space of measurement error rates, $q$, and physical error rates, $p$, where $\mathcal{C}>1$. Figure~\ref{noisy_individual} shows this region for the 9-qubit surface code code, and the 15-qubit gauge color code under depolarising noise. For the surface code, it is possible to demonstrate error suppression with measurement error rates up to $\sim 0.5\%$, but to achieve $\mathcal{C}>2$ measurements must suffer errors at a rate $q<0.1\%$. The other surface codes, and 2D color codes show a similar pattern (see Figure~\ref{noisy_measurements}) with very low measurement error rates needed to demonstrate any significant error suppression. The gauge color code on the other hand, exhibits very different behaviour since it can detect and suppress both $q$ and $p$ type errors. The inset in the figure shows a zoom on the region where $0 < p < 3\%$ and $0 < q < 0.3\%$ which corresponds to the most likely region of experimental interest. The black dotted line indicates the points at which the correcting power of the two codes is equal. Above this line GCC15 outperforms S9.

When the physical noise is asymmetric between the $X$ and $Z$ channels the regions of correctability change. Three cases are shown in Figure~\ref{noisy_measurements}(b)-(d) for the cases of $p'_z = p'_x$, $p'_z = 2 p'_x$ and $p'_z = 5 p'_x$. The symmetric color codes' correctable regions simply shrink as the errors become more unevenly distributed. For asymmetric surface codes, on the other hand, the region of $\mathcal{C}>1$ may grow with the asymmetry in the errors. The 7 qubit surface code, for example, which has no error correcting power under any $(p,q)$ configuration when $p'_z = p'_x$, can suppress errors of up to almost 15\% when $p'_z = 5p'_x$. More details on this are given in Appendix~\ref{appendix:noisy_measurement} where we plot the growth of the region of correctability under varying levels of asymmetry. Given that this type of asymmetry in noise is common in physical systems, this large difference in the error correcting capabilities under such noise models should motivate a choice of code type based on more detailed knowledge about the physical noise of a given system.

\section{Conclusion}

The surface code, color codes, and gauge color code each represent a promising approach to scalable fault tolerance, and we have studied the minimum requirements to implement the most basic instances of each family of codes. The surface code requires a minimum of 9 qubits in order to protect against every single qubit error, but under certain noise patterns 7 and 8-qubit surface codes can suffice to demonstrate error suppression. The color code can be demonstrated with 7 qubits, and the gauge color code with 15 qubits. 

The error model we have considered here is very simplistic, assuming that the only source of noise is decoherence from the environment, so-called memory errors, and allowing perfect preparation and readout of the encoded state. In a physical implementation all operations performed on the qubits will also be faulty. To fully understand the error correcting capabilities of a code in a particular system, one would need to carry out a simulation with the noise model relevant to that system. However, the simple error model we have used here can still give an insight into correcting capabilities, as long as the rates of operational errors are low enough. The additional circuit level noise can be roughly accounted for by increasing the physical error suffered by the qubit. A qubit acquires an  error each time a gate is performed on it. If we consider the example of the 9-qubit circuit code this is between 2-4 times depending on the qubit. In ion trap systems gate errors of $\sim0.1\%$~\cite{Ballance2016} have been demonstrated, and so the additional physical error incurred during stabilizer measurement is $~\sim 0.3\%$. In this case we can identify a modified correcting power,
\[ \mathcal{C}' = \frac{1-(1-p)(1-q)}{p_L(p',q)}, \]
where $p'= p+0.3\%$. Now to argue that the code can demonstrate true error suppression we must show that the modified correcting power is greater than one, $\mathcal{C}'>1$. With a measurement error of $~\sim 0.1\%$~\cite{Harty2014} a memory error of p=5\% can still be effectively suppressed. 

Furthermore, we have demonstrated that the structure of the noise can have a significant impact on the error-correcting capabilities of a code, and choosing an appropriate code structure can increase the correcting power for a given set of physical parameters. Beyond the very simple asymmetries we addressed here, further structure in a noise model is also likely to be reflected in the correcting power. By incorporating the knowledge of the error model into the decoder the correcting capabilities of a code can potentially be boosted even further. 

\bigskip

\begin{acknowledgments}
The author would like to acknowledge Tom Close for the original conception of the precomputed decoder approach used here for analysing the performance of small codes. Further thanks go to Simon Benjamin, Joe O'Gorman and Benjamin Brown for many useful conversations. This research was supported in part by the EPSRC funded Centre for Doctoral Training in Controlled Quantum Dynamics. In preparing and revising the present manuscript, the author acknowledges input from participants in the eQual project, supported by the Office of the Director of National Intelligence (ODNI), Intelligence Advanced Research Projects Activity (IARPA), via the U.S. Army Research Office grant W911NF-16-1-0070. 
\end{acknowledgments}

\bibliography{small_codes}

\appendix

\section{A Precomputed Decoder}
\label{appendix:decoder}

Here we elaborate on the decoding approach used in the main text to determine the optimal error correcting ability of quantum error correcting codes, according to~\cite{TomThesis}. 

We recall that the procedure for error correction involves measurement of the stabilizers of the code, which identifies a syndrome, $s \in S $. The syndrome is some configuration of `+1' and `-1' outcomes, which depend on the error, $E$, that has occurred on the code qubits.  We wish to identify a {\em correction operator}, $C$ - some configuration of Pauli operations that when applied to code will return the state to the code space without resulting in a logical error, such that
\[ C E \ket{\psi_c} = S \ket{\psi_c}, \]
where $\mathcal{S}$ is the stabilizer group, $S \in \mathcal{S}$, and $\ket{\psi_c}$ is the originally encoded state. 

The task of the decoder given a syndrome, $s$, is to identify a correction operator that minimises the chance of acquiring a logical error on the code. Here we describe a method of explicitly calculating the optimal performance of such a decoder for small system sizes.

\subsection{Perfect Measurement}

It is useful to start with an arbitrary correction, $C^*(s)$, for the given syndrome, $s$. Overall the operation $T = C^*E$ has been applied to the code state. This operation is guaranteed to return the code to the code space, but we do not yet know the logical state of the encoded qubit. We can then split the possible physical qubit error configurations, $E \in \mathcal{E}_s$, that could correspond to the observed syndrome, into subsets $\mathcal{E}_{s,l}$ based on the corresponding logical operation, $l\in L$, that is performed by $T$. 

If measurements are error free then we need only consider the rate of physical errors to identify the most probable logical state. Given a syndrome $s$, the probability of introducing the logical operation $l$ on the encoded qubits is, 

\begin{align}
  p(l \vert s) = \sum_{E \in \mathcal{E}_{s,l}} \frac{p(E)}{p(s)}. 
\end{align}

For each  $s \in S$ we identify the most likely logical error $l_s$, such that $p(l_s \vert  s)$ is maximised. The overall successful decoding probability is then given by summing over every possible syndrome,

\begin{align}
  P_d &= \sum_{s \in S}\max_{l\in L}  p(l \vert s)p(s) \\
  &= \sum_{s \in S} \max_{l\in L} \left\{ \sum_{E \in \mathcal{E}_{s,l}} \frac{p(E)}{p(s)} \right\} p(s) \\
  &= \sum_{s \in S} \max_{l\in L} \left\{ \sum_{E \in \mathcal{E}_{s,l}} p(E) \right\}. \label{truthful_prob}
\end{align}

This result is conceptually simple, but computing $P_d$ is challenging due to the rapid growth in the size of the sets $\mathcal{E}_{s,l}$ and  $S$.  With perfect measurements we can make use of the fact that only a restricted set of syndromes are possible to reduce the sum over $s$, but all error configurations must still be considered. Fortunately, as we now show, it is not necessary to compute the the sum over the complete set of error configurations in order to compute $P_d$.  
Obverse that the probability of error configuration $E \in \mathcal{E}$ is 
\begin{align}
  p(E) = (1-p)^{N_Q - n(E)} p^{n(E)},
\end{align}
where $N_Q$ is the number of qubits in the code, and $n(E)$ counts the number of single-qubit errors in $E$. Summing over the errors consistent with a given logical error and syndrome we find 
\begin{align}
  \sum_{E \in \mathcal{E}_{s,l}} p(E) &= \sum_{E \in \mathcal{E}_{s,l}} (1-p)^{N_Q - n(E)} p^{n(E)} \\
  &= (1-p)^{N_Q} \sum_{i = 0}^{N_Q} d_{s,l}^{(i)} \left(\frac{p}{1-p}\right)^i \\
  &=: (1-p)^{N_Q} \chi_{s,l}\left(\frac{p}{1-p}\right),
\end{align}
where $d_{s,l}^{(i)} = \vert \left\{E \in \mathcal{E}_{s,l} : n(E)=i \right\} \vert$ and we have used the final line to define the characteristic function $\chi_{s,l}$ of the class $\mathcal{E}_{s,l}$. By computing and storing the coefficients $d_{s,l}$ we are able to calculate the success probabilities for a range of values of $p$.

\subsection{Imperfect measurements}
Let us now consider the case when in addition to physical errors, stabilizer measurements misreport with probability $q$. When considering noisy measurements we can no longer restrict the set of syndrome outcomes that we sum over, and must instead consider all $2^{N_{S}}$ possibilities for the `true syndrome', $s$, where $N_S$ is the number of stabilizers. We denote the observed syndrome as $s'$. The probability of successful decoding is now given by, 
\begin{align}
P_d &= \sum_{s'\in S}{ P_{s'}p(s') } = \sum_{s'\in S}{\max_{l \in L}{p(l|s')}p(s') },
\label{pd_noisy}
\end{align}
where the sum now runs over all possible syndromes. The probability of a given logical error having occurred must now take into account all possible `true syndromes', 
\begin{align}
p(l|s') = \sum_{s \in S}{p(s|s')p(l|s)}.
\end{align}
Substituting this into \ref{pd_noisy}, we find, 
\begin{align}
P_d &= \sum_{s' \in S}{\max_{l\in L} \sum_{s \in S}{p(s|s')p(s')p(l|s) } }\\
&= \sum_{s' \in S}{\max_{l\in L} \sum_{s \in S}{p(s'|s) p(s)p(l|s) }}\\
&= \sum_{s' \in S}{\max_{l\in L}  \sum_{s \in S}{p(s'|s)p(s)\sum_{E\in \mathcal{E}_{l,s}}{ \frac{p(E)}{p(s)}}}  }\\
&= \sum_{s' \in S}{\max_{l\in L} \sum_{s \in S}{ p(s'|s)\sum_{E\in \mathcal{E}_{l,s}}{p(E)}} }.
\end{align}

Identifying the best choice of matching in this case requires consideration of both the probability of error configurations, $p(E)$, which depends on the physical error rate, $p$, and the probability of the observed syndrome, which depends on the measurement error rate, $q$. The above sum can then be rewritten as in the case of perfect measurement,

\begin{eqnarray}
P_d   & =  \sum_{s' \in S} {\max_{l\in L} } \sum_{s \in S} \ 
& {(1-q)^{N_S - |s-s'|} q^{|s-s'|}}  \\ \nonumber
& & \times \ (1-p)^{N_Q} \chi_{s,l}\left(\frac{p}{1-p}\right)  \\  \nonumber
& = (1-p)^{N_Q}(1-q)^{N_S } \\  \nonumber
& \qquad \times \sum_{s' \in S}  {\max_{l\in L} } \sum_{s \in S} & {\left(\frac{q}{1-q}\right)^{ |s-s'|} \chi_{s,l}\left(\frac{p}{1-p}\right)}. \\  \nonumber
 \label{success_noisy}
\end{eqnarray}
Thus the polynomial functions we computed for the case of perfect measurement can be used also for the case of noisy measurement. Here, however, we cannot restrict the sum to a subset of the functions, but must sum over all with the correct weighting given by Eq.~\ref{success_noisy}.

\subsection{Decoder implementation} 

We implement the decoding strategy outlined in a general form such that the method can be directly applied to any stabilizer code that is defined by its stabilizer generators and logical operators. For a specified code we compute the $\chi_{s,l}$ which can subsequently be used to analyse the logical failure rate of the code under the various noise models we consider in the main text. 

\subsection{Approximating the polynomials}
\begin{figure}
\begin{center}
\includegraphics[width=\columnwidth]{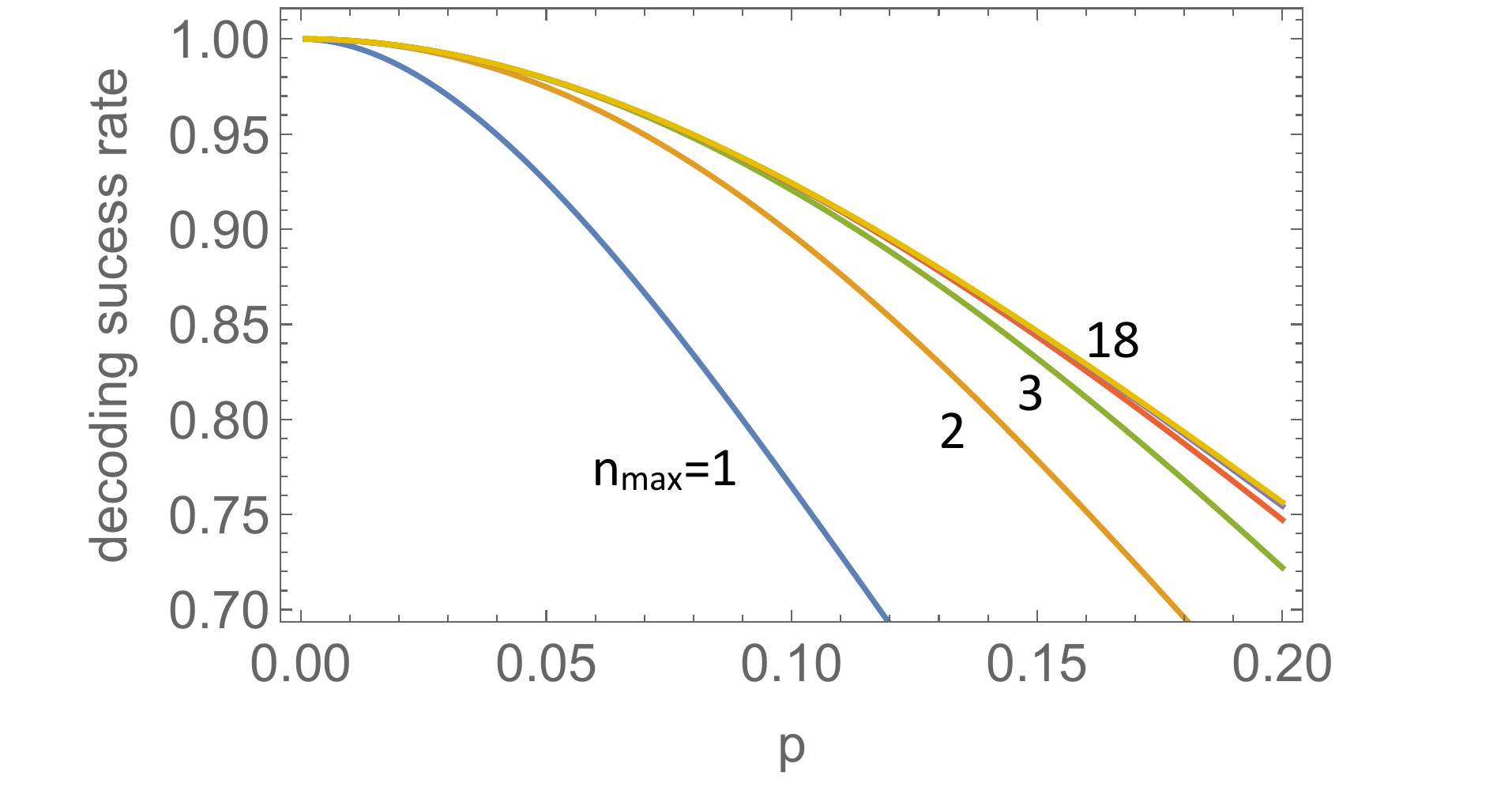}
\end{center}
\caption[Approximating decoding polynomials]{\textbf{Approximating decoding polynomials.} Decoder success rate for the 9 qubit surface code under different levels of truncation approximation. The success rate was calculated for $n_{\mathrm{max}}$ of 1,2,3,4,5 and 18, which is the maximum error rate (each of the 9 qubits can acquire both an $X$ and $Z$ error. For the range of $p<0.2$ the truncated polynomials tend very quickly to the exact value, with the $n_{\mathrm{max}}=5$ case only differing by 0.04\% from the exact result at $p=15\%$.}
\label{approx_poly}
\end{figure}

The computed tables of polynomials provide an exact solution of the decoding success rate. In the small $p$ limit however, many of the higher order terms have a negligible contribution to the overall success probability. To reduce the computational resources required to analyse the decoder in larger codes, and with noisy measurements, we truncate the polynomials at some $n_{\mathrm{max}}$, such that only error configurations containing up to this number of single qubit errors are considered. Our estimate of the success of the decoder is then calculated using the modified polynomial functions,
\[ 
\tilde{\chi}_{s,l}\left(\frac{p}{1-p}\right) =  \sum_{i = 0}^{n_{\rm max}} d_{s,l}^{(i)} \left(\frac{p}{1-p}\right)^i.
\]

We note that this approximation always gives a lower bound on the performance of the decoder, it can never exceed the true value. To avoid assuming some structure in the excluded states, we conservatively treat all the excluded configurations as failures. As an example we show the result of truncation for the 9-qubit surface code in Figure~\ref{approx_poly}. At a physical error rate of 15\% and choosing $n_{\rm max} = 5$ differs from the exact result by only 0.04\%.  

\begin{figure}
\begin{center}
\includegraphics[width=\columnwidth]{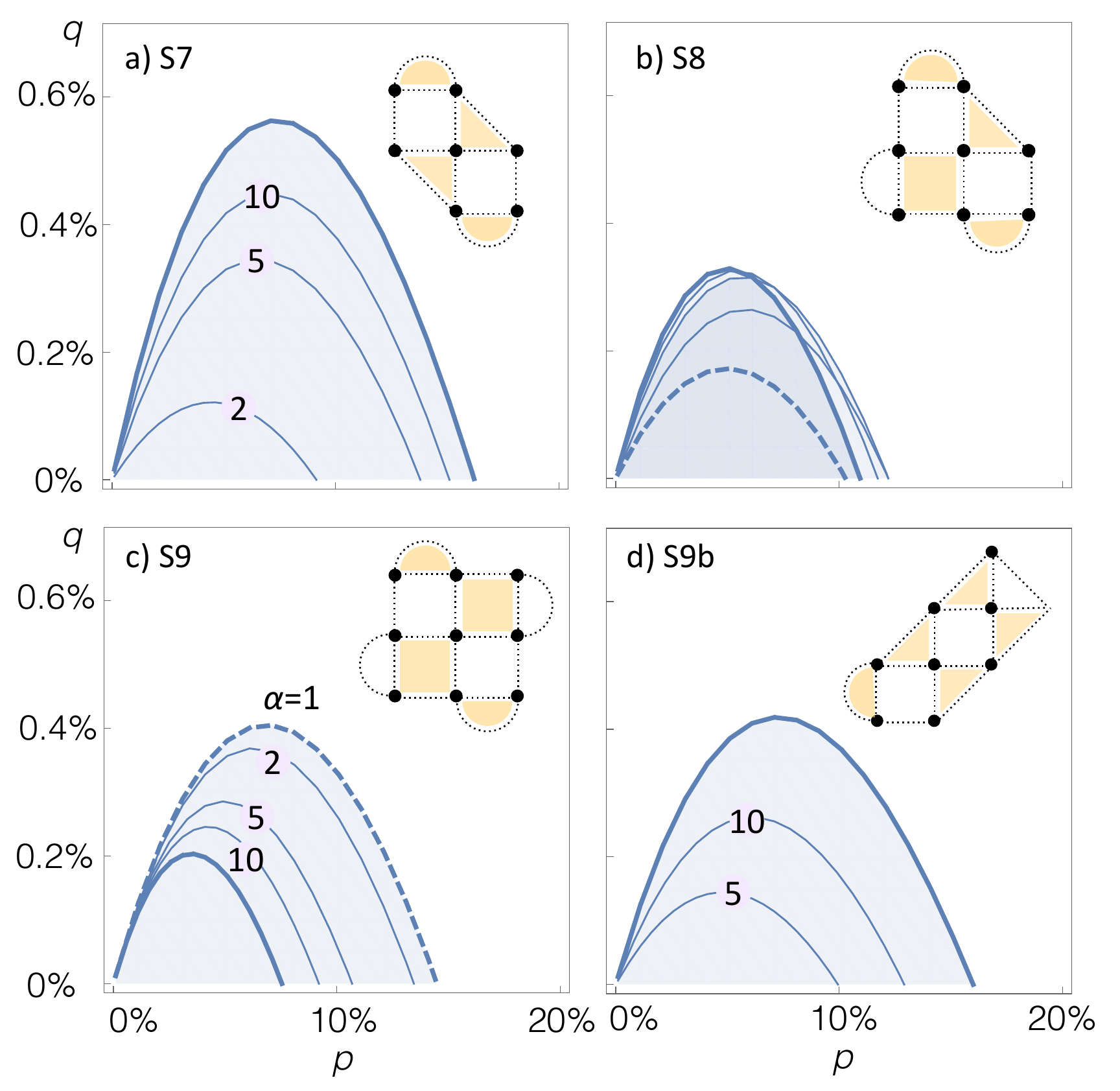}
\end{center}
\caption[Regions of correctability in asymmetric surface codes.]{\textbf{Regions of correctability in surface codes under asymmetric independent noise.} All plots are calculated under an independent noise model where $p'_z = \alpha p'_x$, the contours indicate the border of the region where error suppression is possible for a particular value of $\alpha$. (a) 7 qubit surface code (b) 8 qubit surface code, (c) symmetric 9 qubit surface code and (d) an asymmetric 9 qubit surface code. The code for each plot is shown inset, following the scheme of Figure~\ref{fig:SC_surface_codes}, where orange faces indicate an $X$-type stabilizer and white faces indicate a $Z$-type stabilizer. }
\label{fig:asymmetric}
\end{figure}

\section{Noisy measurement}
\label{appendix:noisy_measurement}
\label{app:SC_asymmetry}

In the main text we saw that an asymmetry in the rate of $X$- and $Z$-type errors could change a code's performance if the stabilizers were not symmetric between the two channels. Figure~\ref{fig:asymmetric} shows how the region of correctability (where $\mathcal{C}>1$) depends on the asymmetry between the two channel, as quantified by parameter $\alpha$, where $p'_z = \alpha p'_x$. The same results hold if $p'_x = \alpha p'_z$, in which case one would invert the $X$- and $Z$- type stabilizers of the codes. $\alpha=1$ represents the symmetric case, and the boundary of the correctable region is denoted by a dashed line in the Figure. $\alpha=\infty$ represents the case where $p'_x = 0$. which is denoted by the solid bold line in the figure. S9 (Fig.~\ref{fig:asymmetric}(c)) is symmetric and so the performance decreases as $\alpha$ increases. S8 shows a moderate increase in the $q$ direction with increased $\alpha$. The most dramatic increase in the correctable region is seen in S7 and S9b, both have no correcting power for $\alpha=1$, but increasing to $\alpha=5$ both can demonstrate error suppression at physical error rates of over 10\%.

\end{document}